\documentclass[fleqn,11pt]{article}

\usepackage{amsmath,amsfonts,amssymb}
\usepackage{color}

\usepackage{geometry}
\newtheorem{theorem}{Theorem}
\newtheorem{lemma}[theorem]{Lemma}

\newtheorem{definition}{Definition}

\newcommand{\R}{\mathbb{R}}

\newcommand{\N}{\mathbb{N}}

\newcommand{\cqfd}
{%
\mbox{}%
\nolinebreak%
\hfill%
\rule{2mm}{2mm}%
\newline
\newline
}

\title{On the Cauchy problem with large data for the space-dependent Boltzmann Nordheim equation III.}
\author{Leif ARKERYD and Anne NOURI\\
\\Mathematical Sciences, 41296 G\"oteborg, Sweden,\\
arkeryd@chalmers.se\\
Aix-Marseille University, CNRS, Centrale Marseille, I2M UMR 7373, 13453 Marseille, France,\\
anne.nouri@univ-amu.fr}

\date{}

\begin{document}

\maketitle

{\noindent \bf Abstract.}\hspace{0.1in}
This paper studies the quantum Boltzmann Nordheim equation from a Boltzmann equation for Haldane statistics. Strong  solutions are obtained for the Cauchy problem with large initial data in an $L^1\cap L^\infty$ setting on a torus $[0,1]$ (resp. $[0,1]^k$, $k\in \{ 2,3\}$) for three-dimensional velocities and pseudo-Maxwellian (resp. very soft) forces. The main results are existence, uniqueness and stability of solutions conserving mass, momentum and energy that explode in $L^\infty $ if they are only local in time.\\

\footnotetext[2]{Key words; bosonic Boltzmann-Nordheim equation, low temperature kinetic theory, quantum Boltzmann equation.}
%
%
%
%
\section{Introduction and main result.}
The quantum Boltzmann Nordheim equation for bosons was first derived by Nordheim \cite{N}, Uehling and Uhlenbeck \cite{UU}. In a one-dimensional periodic spatial and three-dimensional in velocity setting, it is
\begin{equation}\label{f}
\partial _tf(t,x,v)+v_1\partial _xf(t,x,v)= Q(f)(t,x,v),\quad t>0, \quad x\in [0,1],\quad v= (v_1,v_2,v_3)\in \R ^3,
\end{equation}
where
\begin{equation}\label{Q}
Q(f)(v)= \int_{I\! \!R^3 \times \mathcal{S}^2}B
 [f^\prime f^\prime _*F(f)F(f_*)-ff_*F(f^\prime )F(f^\prime _*)] dv_*dn ,\hspace{.1cm}
\end{equation}
$B$ is a given collision kernel,
\begin{align*}
&v^\prime = v-(v-v_*,n)n, \hspace*{0.03in}v^\prime _*= v_*+(v-v_*,n)n, \hspace*{0.03in}f^\prime = f(v^\prime),\hspace*{0.03in}f^\prime_*= f(v^\prime _*), \hspace*{0.03in}f_*= f(v_*),
\end{align*}
and
\begin{equation}\label{F}
F(f)= 1+f.
\end{equation}
For the Boltzmann Nordheim equation, general existence results were first obtained by X. Lu in \cite{Lu1} in the space-homogeneous isotropic large data case. It was followed by a number of interesting studies in the same isotropic setting, by X. Lu \cite{Lu2,Lu3,Lu4}, and by M. Escobedo and J.L. Vel\'azquez \cite{EV2,EV}. Space-homogeneous results with the isotropy assumption removed, were obtained by M. Briant and A. Einav \cite{BE}. Finally a space-dependent case close to equilibrium has been studied by G. Royat in \cite{R}.\\
The papers \cite{Lu1,Lu2,Lu3,Lu4} by Lu, study the isotropic, space-homogeneous Boltzmann Nordheim equation both for Cauchy data leading to mass and energy conservation, and for data leading to mass loss when time tends to infinity. Escobedo and Vel\'asquez in \cite{EV2,EV}, again in the isotropic space-homogeneous case, study initial data leading to concentration phenomena and blow-up in finite time of the $L^\infty$-norm of the solutions. The paper \cite{BE}  by Briant and Einav removes the isotropy restriction and obtains existence and uniqueness on a time interval $[0,T_0)$ in polynomially weighted spaces of $L^1\cap L^\infty$ type. In \cite{BE} either $T_0=\infty$, or for finite $T_0$ the $L^\infty$-norm of the solution tends to infinity when time tends to $T_0$. Finally the paper \cite{R} considers the space-dependent problem, for a particular setting close to equilibrium, and proves well-posedness and convergence to equilibrium. \\
The analysis in the present paper is based on the results related to the Boltzmann equation for Haldane statistics in \cite{AN1}, by a limiting procedure when $\alpha\rightarrow 0$.\\
In a previous paper \cite{AN1}, we have solved the Cauchy problem for a space-dependent Boltzmann equation for Haldane statistics,
\begin{align}\label{f-alpha}
&\partial _tf(t,x,v)+\bar{v}\cdot \nabla _xf(t,x,v)= Q_{\alpha }(f)(t,x,v)
,\nonumber \\
&\hspace*{1.in} {(t,x)\in \R _+\times [ 0,1] ^k,\hspace*{0.02in} v=(v_1,v_2,v_3)\in \R ^3}, \hspace*{0.02in}\bar{v}= (v_1,\cdot \cdot \cdot ,v_k), \hspace*{0.02in}k\in \{ 1, 2,3\} .
\end{align}
The collision operator $Q_{\alpha }$ in \cite{AN1} depends on a parameter $\alpha \in ] 0,1[ $ and is given by
\begin{eqnarray*}
Q_\alpha (f)(v)= \int_{I\! \!R^2 \times \mathcal{S}^2}B
 [f^\prime f^\prime _*F_\alpha(f)F_\alpha(f_*)-ff_*F_\alpha(f^\prime )F_\alpha(f^\prime _*)] dv_*dn,\hspace{.1cm}
\end{eqnarray*}
with the filling factor $F_\alpha$
\begin{eqnarray*}
F_\alpha(f)= (1-\alpha f)^{\alpha}(1+(1-\alpha)f)^{1-\alpha}\, .
\end{eqnarray*}
The  study of quantum elementary and quasi-particles at low temperatures including their statistics, is a frontier area in modern physics. The definition proposed by Haldane in a seminal paper \cite{H} is based on a generalization of the Pauli exclusion principle for fractional quantum statistics valid for any dimension. The conventional Bose-Einstein and Fermi-Dirac statistics are commonly associated with integer spin bosonic elementary particles resp. half integer spin fermionic elementary  particles, whereas the Haldane fractional statistics  is connected with quasi-particles corresponding to elementary excitations in many-body interacting quantum systems.\\
In \cite{AN1}, also the limiting case $\alpha=1$ is discussed, the Fermi-Dirac equation \cite{N} for fermions.
In the present paper we consider the other limiting case,  $\alpha=0$, which is the Boltzmann Nordheim equation for bosons. \\
\\
Denote by
\begin{equation}\label{f-sharp}
f^{\sharp }(t,x,v)= f(t,x+tv_1,v)\quad (t,x,v)\in \R _+\times [ 0,1] \times \R ^3.
\end{equation}
Strong solutions to the Boltzmann Nordheim equation are considered in the following sense.
\begin{definition}\label{strong-solution}
$f$ is a strong solution to (\ref{f}) on the time interval $I$ if
\begin{eqnarray*}
f\in\mathcal{C}^1(I;L^1([0,1]\times\R^3)),
\end{eqnarray*}
and
\begin{equation}\label{eq-along-characteristics}
\frac{d}{dt}f^{\sharp }= \big( Q(f)\big) ^{\sharp },\quad \text{on   } I\times [ 0,1] \times \R ^3.
\end{equation}
\end{definition}
With
\begin{eqnarray*}
\cos \hspace*{0.02in} \theta = n \cdot \frac{v-v_*}{|v-v_*|}\, ,
\end{eqnarray*}
the kernel $B(|v-v_*|, \theta )$ is assumed measurable with
\begin{equation}\label{hyp1-B}
0\leq B\leq B_0,
\end{equation}
for some $B_0>0$. It is also assumed for some $\gamma, \gamma',{c_B}>0$, that
\begin{equation}\label{hyp2-B}
B(|v-v_*|, \theta )=0 \hspace*{0.05in}\text{for}\hspace*{0.05in}   |\cos \hspace*{0.02in}\theta |<\gamma',\quad
\text{for}\hspace*{0.05in}  1-|\cos \hspace*{0.02in}\theta|<\gamma',\quad  \text{and for   } |v-v_*|< \gamma .
\end{equation}
%
%
The main result of this paper is the following.
\begin{theorem}\label{main-theorem}
Assume (\ref{hyp1-B})-(\ref{hyp2-B}). Let $f_0\in L^\infty([0,1]\times\R^3)$ and satisfy
\begin{align}
&(1+|v|^2)f_0(x,v) \in L^1([ 0,1] \times \R ^3), \label{hyp-f0-1}\\
&\int \sup_{x\in[0,1]} f_0(x,v)dv=c_0 <\infty . \label{hyp-f0-2}
\end{align}
There exist a time $T_\infty >0$ and a strong solution $f$ to (\ref{f}) on $[0,T_\infty)$ with initial value $f_0$. \\
For $0<T<T_\infty$, it holds
\begin{equation}\label{regularity-f}
f\in \mathcal{C}^1([0,T_\infty);L^1([0,1]\times\R^3))\cap L^\infty ([0,T]\times[0,1]\times \R^3).
\end{equation}
If $T_\infty <+\infty $ then
\begin{equation}\label{explosion}
\overline{\lim}_{t\rightarrow T_\infty }\parallel f(t,\cdot ,\cdot )\parallel _{L^\infty ([ 0,1] \times \R ^3)}= +\infty .
\end{equation}
The solution is unique, and conserves mass, momentum, and energy. For equibounded families in $L^\infty ([ 0,1] \times \R ^3)$ of initial values, the solution depends continuously in $L^1$ on the initial value $f_0$.
\end{theorem}
\underline{Remarks.}\\
The strong cut-off conditions on $B$ are made for mathematical reasons. For a more general discussion of cut-offs in the collision kernel $B$, see \cite{Lu2}. \\
A result  analogous to Theorem \ref{main-theorem} has been proven in the slab and two-dimensional velocities \cite{AN2} from the resolution of the Cauchy problem for anyons performed in  \cite{AN3}. There, the two-dimensional velocities frame is basic both for the physics setting of anyons and the mathematical proofs.\\
\hspace*{1.in}\\
A result similar to Theorem \ref{main-theorem} holds in any dimension for the physical space, under a supplementary assumption on the collision kernel $B$. Consider the problem
\begin{align}
&\partial _tf(t,x,v)+\bar{v}\cdot \nabla _xf(t,x,v)= Q(f)(t,x,v),\hspace*{0.03in}(t, x, v)\in [0,T] \times [ 0,1] ^k\times \R ^3 ,\hspace*{0.03in}  v=(v_1,v_2,v_3) \in \R ^3, \label{f-soft-potentials} \\
&f(0,x,v)= f_0(x,v),\label{init}
\end{align}
where
\begin{align*}
&\bar{v}= (v_1)\hspace*{0.03in} (\text{  resp.  } \bar{v}= (v_1, v_2), \text{  resp.  } \bar{v}= v) \text{  for  }k=1\hspace*{0.03in} (\text{  resp.  } k=2, \text{  resp.  }k= 3).
\end{align*}
The following result holds with an analogous proof to the one of Theorem \ref{main-theorem}.
\begin{theorem}\label{th-global}
\hspace*{0.1in}\\
Under the assumptions of Theorem \ref{main-theorem} and the supplementary assumption of very soft collision kernels,
 \begin{equation}\label{hyp4-B}
B(u, \theta )= B_1(u)B_2(\theta )\quad \text{with   }|B_1(u)|\leq c|u|^{-3-\eta }\text{   for some  }\eta >0, \text{and  }B_2\text{   bounded},
\end{equation}
the statement of Theorem \ref{main-theorem} holds for  (\ref{f-soft-potentials})-(\ref{init}).
\end{theorem}
\hspace*{1.in}\\
To obtain Theorem 1.1, we start from a fixed initial value $f_0$ bounded by $2^L$ with $L\in \N$. We prove that there is
a time $T_L>0$ independent of $\alpha\in ] 0,2^{-L-1}[ $, such that a family of approximations $(g_\alpha )$ are bounded by $2^{L+1}$ on $[0,T_L]$. We then prove that the limit $f$ of  $(g_\alpha )$ when $\alpha\rightarrow 0$  solves the  Boltzmann Nordheim problem.
Iterating the result from $T_L$ on, it follows that $f$ exists up to the first time $T_\infty$ when $ \overline{\lim }_{t\rightarrow T_\infty}  \parallel f(t,\cdot ,\cdot )\parallel _{L^{\infty }([0,1] \times \R ^3)} = +\infty $.\\
The paper is organized as follows. In the following section, approximations $(g_\alpha )$ to the Cauchy problem for a Boltzmann equation for Haldane statistics slightly modified with respect to \cite{AN1} are constructed. Their mass density are locally in time uniformly controlled with respect to $\alpha $ with the help of their Bony functionals. Theorem \ref{main-theorem} is proven in Section 3 except for the conservations of mass, momentum and energy that are proven in Section 4.
\hspace*{0.1in}\\
%
%
%
%
\section{Approximations}
\setcounter{equation}{0}
\setcounter{theorem}{0}
For $\alpha \in ] 0,1[ $, denote by $\psi_\alpha $ the cut-off function with
\[ \begin{aligned}
&\psi_\alpha (r)=0\quad \text{if   }r>\frac{1}{\alpha ^2}
&\text{and}\quad \psi_\alpha (r)= 1\quad \text{if  }
r\leq \frac{1}{\alpha ^2},
\end{aligned}\]
and set
\[ \begin{aligned}\chi_\alpha (v,v_*)=\psi_\alpha (|v ^2| +|v_*| ^2 ) .
\end{aligned}\]
Let $G_\alpha $ be the $C^1$ function defined on $[ 0,\frac{1}{\alpha }] $ by
\begin{eqnarray*}
 G_\alpha (y)= \frac{1-\alpha y}{(\alpha +1-\alpha y)^{1-\alpha }} (1+(1-\alpha )y)^{1-\alpha }.
\end{eqnarray*}
Denote by $\tilde{Q}_\alpha $ (resp. $\tilde{Q}_\alpha ^+$, {and $\tilde{Q}_\alpha ^-$ to be used later}), the operator
\begin{align}
&\tilde{Q}_\alpha (f)(v):= \int B(|v-v_*|,{\theta} )\chi _\alpha (v,v_*)\Big( f^\prime f^\prime _*G_\alpha (f)G_\alpha (f_*)-ff_*G_\alpha (f^\prime )G_\alpha (f^\prime _*)\Big) dv_*dn  ,\nonumber \\
&(\text{resp. its gain part    }\tilde{Q}_\alpha ^+(f)(v):= \int B(|v-v_*|,{\theta } )\chi _\alpha (v,v_*)f^\prime f^\prime _*G_\alpha (f)G_\alpha (f_*)dv_*dn ,\label{Qtildealpha+} \\
&{\text{and its loss part    }\tilde{Q}_\alpha ^-(f)(v):= \int B(|v-v_*|,{\theta } )\chi _\alpha (v,v_*)ff_*G_\alpha (f^\prime )G_\alpha (f^\prime _*)dv_*dn \hspace*{0.02in} ).}\label{Qtildealpha-}
\end{align}
Let a mollifier $\varphi _\alpha $ be defined by  $\varphi _\alpha (x,v)= \frac{1}{\alpha ^4}\varphi (\frac{x}{\alpha },\frac{v}{\alpha })$, where
\begin{align*}
&\varphi \in C_0^\infty (\R ^{4}),\quad support (\varphi )\subset [ 0,1] \times \{ v\in \R ^3;\lvert v\rvert \leq 1\},\quad \varphi \geq 0,\quad\int _{[0,1] \times \R ^3}\varphi (x,v)dxdv= 1.
\end{align*}
Let
\begin{equation}\label{df-init-approx}
f_{0,\alpha } \text{  be the restriction to  }[ 0,1] \times \{v; \lvert v\rvert \leq \frac{1}{\alpha }\} \text{  of   }\big( \min \{f_0, \frac{1}{\alpha }-\alpha \} \big) \ast \varphi _\alpha .
\end{equation}
The following lemma concerns approximations for  $(x,v)\in [0,1] \times \{ v\in \R ^3; \lvert v\rvert \leq \frac{1}{\alpha }\}$.\\
%
%
\begin{lemma}\label{approximation-alpha}
\hspace{1cm}\\
For any $T>0$, there is a unique solution $g_\alpha \in {C}\big( [ 0,T] \times [ 0,1] ;L^1( \{v; \lvert v\rvert \leq \frac{1}{\alpha }\} )\big) $ to
\begin{equation}\label{eq-g-alpha}
\partial _tg_\alpha +v_1\partial _xg_\alpha = \tilde{Q}_\alpha (g_\alpha ),\quad g_\alpha (0,\cdot ,\cdot )= f_{0,\alpha }.
\end{equation}
There is $\eta _\alpha >0$ such that $g_\alpha $ takes its values in $ [ 0,\frac{1}{\alpha }-\eta _\alpha ] $.\\
The solution conserves mass, momentum and energy.
\end{lemma}
\underline{Proof of Lemma \ref{approximation-alpha}.}\\
Let $T>0$ be given. We shall first prove by contraction that for $T_1>0$ and small enough, there is a unique solution
\begin{eqnarray*}
g_\alpha \in C\big( [ 0,T_1] \times [ 0,1] ; L^1( \{ v; \lvert v\rvert \leq \frac{1}{\alpha }\} )\big) \cap \{f; f\in [ 0,\frac{1}{\alpha } ] \}
\end{eqnarray*}
to (\ref{eq-g-alpha}). Let the map $\mathcal{C}$ be defined on {periodic in $x$ functions in}
\begin{eqnarray*}
C \big( [ 0,T] \times [ 0,1] ;L^1( \{ v; \lvert v\rvert \leq \frac{1}{\alpha }\} )\big) \cap \{f; f\in [ 0,\frac{1}{\alpha } ] \}
\end{eqnarray*}
by $\mathcal{C}(f)= g$, where
\[ \begin{aligned}
&\partial _tg +v_1\partial _xg = (1-\alpha g)\Big( \frac{1+(1-\alpha )f}{\alpha +1-\alpha f}\Big) ^{1-\alpha }\int B\chi _\alpha{f}^\prime {f}^\prime _*G_\alpha ({f}_*)dv_*dn -g\int B\chi _\alpha {f}_*G_\alpha (f^\prime )G_\alpha (f^\prime _*)dv_*dn  ,\\
&g(0,\cdot ,\cdot )= f_{0,\alpha }.
\end{aligned}\]
The previous linear partial differential equation has a unique periodic solution
\begin{eqnarray*}
g\in C \Big( [ 0,T] \times [ 0,1] ;L^1\big( \{ v; \lvert v\rvert \leq \frac{1}{\alpha }\} \big) \Big) .
\end{eqnarray*}
For $f$ with values in $ [ 0,\frac{1}{\alpha } ] $, $g$ takes its values in $ [ 0,\frac{1}{\alpha } ] $. Indeed, $g^\sharp $ defined in (\ref{f-sharp}) satisfies
\[ \begin{aligned}
g^\sharp (t,x,v)&= f_{0,\alpha }(x,v)e^{-\int _0^t\bar{\sigma }_f^\sharp (r,x,v)dr}\\
&+\int_0^tds\Big( \big( \frac{1+(1-\alpha )f}{\alpha +1-\alpha f}\big) ^{1-\alpha }\int B\chi _\alpha {f}^\prime {f}^\prime _*G_\alpha ({f}_*)dv_*dn\Big)^\sharp (s,x,v)e^{-\int _s^t\bar{\sigma }_f^\sharp (r,x,v)dr}\\
&\geq f_{0,\alpha }(x,v)e^{-\int _0^t\bar{\sigma }_f^\sharp (r,x,v)dr}\geq 0,
\end{aligned}\]
and
\[ \begin{aligned}
(1-\alpha g)^\sharp (t,x,v)&= (1-\alpha f_{0,\alpha })(x,v)e^{-\int _0^t\bar{\sigma }_f^\sharp (r,x,v)dr}\\
&+\int _0^t\Big( \int B\chi _\alpha f_*G_\alpha (f^\prime)G_\alpha (f^\prime _*)dv_*dn\Big) ^\sharp (s,x,v)e^{-\int _s^t\bar{\sigma }_f^\sharp (r,x,v)dr}ds\\
&\geq (1-\alpha f_{0,\alpha })(x,v)e^{-\int _0^t\bar{\sigma }_f^\sharp (r,x,v)dr}\geq 0.
\end{aligned}\]
Here,
\[ \begin{aligned}
&\bar{\sigma }_f:= \alpha \Big( \frac{1+(1-\alpha )f)}{\alpha +1-\alpha f}\Big) ^{1-\alpha }\int B\chi _\alpha f^\prime f^\prime _*G_\alpha (f_*)dv_*dn + \int B\chi _\alpha f_*{G}_{\alpha }(f^\prime ){G}_{\alpha }(f^\prime _*)dv_*dn  .
\end{aligned}\]
$\mathcal{C}$ is a contraction on $C\Big( [0,T_1] \times [ 0,1] ; L^1\big (\{ v;\lvert v\rvert \leq \frac{1}{\alpha }\} \big) \Big) \cap \{f; f\in [ 0,\frac{1}{\alpha } ] \}$, for $T_1>0$ small enough only depending on $\alpha $, since the derivative of the map $G_\alpha $ is bounded on $[ 0,\frac{1}{\alpha }] $. Let $g_\alpha $ be its fixed point, i.e. the solution of (\ref{eq-g-alpha}) on $[ 0,T_1] $. The argument can be repeated and the solution continued up to $t=T$. By Duhamel's form for $g_\alpha $ (resp. $1-\alpha g_\alpha $),
\[ \begin{aligned}
g_\alpha ^\sharp (t,x,v)&\geq f_{0,\alpha }(x,v)e^{-\int _0^t\bar{\sigma }_{g_\alpha }^\sharp (r,x,v)dr}\geq 0,
\quad (t, x)\in [ 0,T] \times [ 0,1] ,\hspace{0.03in}\lvert v\rvert \leq \frac{1}{\alpha },
\end{aligned}\]
(resp.
\[ \begin{aligned}
(1-\alpha g_\alpha )^\sharp (t,x,v)&\geq (1-\alpha f_{0,\alpha })(x,v)e^{-\int _0^t\bar{\sigma }_{g_\alpha }^\sharp (r,x,v)dr}\\
&\geq \alpha e^{-c\alpha ^4T},
\quad (t, x)\in [ 0,T]\times [ 0,1] ,\hspace{0.03in}\lvert v\rvert \leq \frac{1}{\alpha }).
\end{aligned}\]
Consequently, for some $\eta_\alpha >0$, there  is a {periodic in $x$} solution
\begin{eqnarray*}
g_\alpha \in C([ 0,T] \times [ 0,1] ;L^1( \{ v;\lvert v\rvert \leq \frac{1}{\alpha }\} ))
\end{eqnarray*}
to (\ref{eq-g-alpha}) with values in $[ 0, \frac{1}{\alpha }-{\eta_\alpha }]$.\\
If there were another nonnegative local solution $\tilde{g}_{\alpha }$ to (\ref{eq-g-alpha}), defined on $[ 0,T^\prime ] $ for some $T^\prime \in ] 0,T] $, then by the exponential form it would stay below $\frac{1}{\alpha}$. The difference $g_\alpha -\tilde{g}_\alpha $ would for some constant $c_{T^\prime}$ satisfy
\begin{eqnarray*}
\int \lvert (g_\alpha  -\tilde{g}_\alpha )^\sharp (t,x,v)\rvert dxdv\leq c_{T^\prime }\int _0^t\int \lvert (g_\alpha -\tilde{g}_\alpha )^\sharp (s,x,v)\rvert dsdxdv,\hspace*{0.03in}t\in [ 0,T^\prime ],\quad  (g_\alpha -\tilde{g}_\alpha )^\sharp (0,x,v)= 0,
\end{eqnarray*}
implying that the difference {would be} identically zero on $[ 0,T^\prime ] $. Thus $g_\alpha $ is the unique solution on $[ 0,T] $ to (\ref{eq-g-alpha}), and has its range contained in $[ 0,\frac{1}{\alpha }-\eta _\alpha ] $. \\
The conservations of mass (resp. momentum, resp. energy) of $g_\alpha $ follow from a direct integration w.r.t. $v$ of (\ref{eq-g-alpha}) (resp. (\ref{eq-g-alpha}) multiplied by $v$, resp (\ref{eq-g-alpha}) multiplied by $\lvert v\rvert ^2$).
 \cqfd
%
%
%
There is a Bony type inequality for $(g_\alpha )$, as follows.
\begin{lemma}\label{bony}
\hspace*{0.1in}\\
Denote by $n_1$ the component along the $x$-axis of $n\in \mathcal{S}^2$, and $g_\infty $ an $L^\infty $ bound of $g_\alpha $ on \\
$[0,T] \times [0,1] \times \R ^3$. It holds that
\begin{align}
&\int_0^T\int n_1^2[(v-v_*)\cdot n]^2B{\chi}_\alpha g_\alpha g_{\alpha *}G_\alpha (g^\prime _\alpha )G_\alpha (g^\prime _{\alpha *})dvdv_*dn dxds\leq c'_0(g_\infty +1)^2(1+T),\nonumber \\
&\hspace*{4.95in} \alpha \in ]0,1[,\label{ineq-bony}
\end{align}
with $c'_0$ only depending on $\int f_0(x,v)dxdv$ and $\int |v|^2f_0(x,v)dxdv$.
\end{lemma}
\underline{Proof of Lemma \ref{bony}.} \\
Denote $g_\alpha $ by $g$ for simplicity.
The integral over time of the first component of momentum $\int v_1g(t,0,v)dv$ (resp. $\int v_1^2g(t,0,v)dv$ ) is first controlled. Let $\beta \in C^1([ 0,1] )$ be such that $\beta (0)= -1$ and $\beta (1)= 1$. Multiply (\ref{eq-g-alpha}) by $\beta (x)$ (resp. $v_1\beta (x)$ ) and integrate over $[ 0,T] \times [ 0,1] \times \R ^3$. It gives
\[ \begin{aligned}
\int _0^T\int v_1g(\tau ,0,v)dvd\tau = \frac{1}{2}\big( \int \beta (x)f_{0,\alpha }(x,v)dxdv&-\int \beta (x)g(T,x,v)dxdv\\
&+\int _0^T\int \beta ^\prime (x)v_1g(\tau ,x,v)dxdvd\tau\big) ,
\end{aligned}\]
\Big( resp.
\[ \begin{aligned}
\int _0^T\int v_1^2g(\tau ,0,v)dvd\tau = \frac{1}{2}\big( \int \beta (x)v_1f_{0,\alpha }(x,v)dxdv&-\int \beta (x)v_1g(T,x,v)dxdv\\
&+\int _0^T\int \beta ^\prime (x)v_1^2g(\tau ,x,v)dxdvd\tau\big) \Big) .
\end{aligned}\]
Consequently, using the conservation of mass and energy of $g$,
\begin{align}\label{bony-1}
\lvert \int _0^T\int v_1g(\tau ,0,v)dvd\tau \rvert +\int _0^T\int v_1^2g(\tau ,0,v)dvd\tau \leq c(1+T).
\end{align}
Let
\begin{eqnarray*}
\mathcal{I}(t)= \int _{x<y}(v_1-v_{*1})g(t,x,v)g(t,y,v_*)dxdydvdv_*.
\end{eqnarray*}
It results from
\begin{eqnarray*}
\mathcal{I}'(t)= -\int (v_1-v_{*1})^2g(t,x,v)g(t,x,v_*)dxdvdv_*+2\int v_{*1}(v_{*1}-v_1)g(t,0,v_*)g(t,x,v)dxdvdv_*,
\end{eqnarray*}
and the conservations of the mass, momentum and energy of $g$ that
\begin{align}\label{bony-2}
&\int _0^T \int_0^1 \int (v_1-v_{*1})^2 g(s,x,v)g(s,x,v_*)dvdv_*dxds\nonumber \\
&\leq 2\int f_0(x,v)dxdv\int \lvert v_1\rvert f_{0,\alpha }(x,v)dv+ 2\int g(T,x,v)dxdv\int \lvert v_1\rvert g(T,x,v)dxdv\nonumber \\
&+2\int _0^T\int v_{*1}(v_{*1}-v_1)g(\tau ,0,v_*)g(\tau ,x,v)dxdvdv_*d\tau \nonumber \\
&\leq 2\int f_0(x,v)dxdv\int (1+\lvert v\rvert ^2)f_0(x,v)dv+ 2\int g(T,x,v)dxdv\int (1+\lvert v\rvert ^2) g(T,x,v)dxdv\nonumber \\
&+2\int _0^t(\int v_{*1}^2g(\tau ,0,v_*)dv_*)d\tau\int f_0(x,v)dxdv-2\int _0^T(\int v_{*1}g(\tau ,0,v_*)dv_*)d\tau\int  v_1f_0(x,v)dxdv\nonumber \\
&\leq c\Big( 1+\int _0^T\int v_1^2g(\tau ,0,v)dvd\tau +\lvert \int _0^T\int v_1g(\tau ,0,v)dv\rvert \Big) . \nonumber
\end{align}
And so, by (\ref{bony-1}),
\begin{equation}\label{bony-3}
\int _0^T \int_0^1 \int (v_1-v_{*1})^2 g(\tau ,x,v)g(\tau ,x,v_*)dxdvdv_*d\tau \leq c(1+T).
\end{equation}
Here, $c$ is a constant depending only on $\int f_0(x,v)dxdv$ and $\int \lvert v\rvert ^2f_0(x,v)dxdv$. \\
Denote by $u_1=\frac{\int v_1gdv}{\int gdv}$. It holds
\begin{align}\label{bony-4}
\int_0^T\int_0^1 \int (v_1-u_1)^2 B{\chi}_\alpha gg_*&G_\alpha (g^\prime )G_\alpha (g^\prime _*)(s,x,v,v_*,n)dvdv_*dn dxds\nonumber \\
&\leq c(g_\infty +1)^2\int_0^T  \int_0^1 \int (v_1-u_1)^2 gg_*(s,x,v,v_*)dvdv_* dxds\nonumber \\
&= \frac{c}{2}(g_\infty +1)^2\int _0^T \int_0^1 \int (v_1-v_{*1})^2 gg_*(s,x,v,v_*)dvdv_*dxds\nonumber \\
&\leq c(g_\infty +1)^2(1+T).
\end{align}
Multiply equation (\ref{eq-g-alpha}) for $g$  by $v_1^2$, integrate and use that $\int v_1^2\tilde{Q}_\alpha (g)dv= \int (v_1-u_1)^2\tilde{Q}_\alpha (g)dv$ and (\ref{bony-4}). It results
\[ \begin{aligned}
&\int _0^T\int (v_1-u_1)^2B{\chi}_\alpha g^\prime g^\prime _*G_\alpha (g)G_\alpha (g_*)dvdv_*dn dxds\\
&= \int v_1^2g(T,x,v)dxdv-\int v_1^2f_{0,\alpha }(x,v)dxdv+\int _0^T\int (v_1-u_1)^2B{\chi}_\alpha gg_*G_\alpha (g^\prime )G_\alpha (g^\prime _*)dxdvdv_*dn ds\\
&\leq c^\prime (g_\infty +1)^2(1+T),
\end{aligned}\]
where $c^\prime $ is a constant only depending on $\int f_0(x,v)dxdv$ and $\int \lvert v\rvert ^2f_0(x,v)dxdv$.\\
\hspace{1cm}\\
After  a change of variables the left hand side can be written
\[ \begin{aligned}
&\int _0^T\int (v'_1-u_1)^2B{\chi}_\alpha gg_*G_\alpha (f^\prime )G_\alpha (g^\prime_*)dvdv_*dn dxds\\
&= \int _0^T\int (c_1-n_1[(v-v_*)\cdot n])^2B{\chi}_\alpha gg_*G_\alpha (g^\prime )G_\alpha (g^\prime _*)dvdv_*dn dxds,
\end{aligned}\]
where $c_1=v_1-u_1$.
Expand $(c_1-n_1[(v-v_*)\cdot n])^2$, remove the  positive term containing $c_1^2$. \\
\\
The term containing $n_1^2[(v-v_*)\cdot n]^2$ is estimated as follows;
\hspace*{0.1in}\\
\begin{align*}
&\int_0^T\int n_1^2[(v-v_*)\cdot n]^2B{\chi}_\alpha gg_*G_\alpha (g^\prime )G_\alpha (g^\prime _*)dvdv_*dn dxds\\
&\leq c^\prime {(1+T)}(g_\infty +1)^2+2\int _0^T\int (v_1-u_1)n_1[(v-v_*)\cdot n]B{\chi}_\alpha gg_*G_\alpha (g^\prime )G_\alpha (g^\prime _*)dvdv_*dn dxds\\
&\leq {c^\prime }{(1+T}(g_\infty +1)^2+2\int _0^T\int \Big( v_1\sum^3_{l=2}(v_l-v_{*l})n_1n_l
\Big) B{\chi}_\alpha gg_*G_\alpha (g^\prime )G_\alpha (g^\prime _*)dvdv_*dn dxds,
\end{align*}
since
\[ \begin{aligned}
\int u_1(v_l-v_{*l})n_1n_l{\chi}_\alpha Bgg_*G_\alpha (g^\prime )G_\alpha (g^\prime _*)dvdv_*dn dx
= \hspace*{0.01in}0,\quad\quad l=2, 3,
\end{aligned}\]
by an exchange of the variables $v$ and $v_*$. Moreover, exchanging first the variables $v$ and $v_*$,
\[ \begin{aligned}
2\int _0^T&\int v_1\sum^3_{l=2}(v_l-v_{*l})n_1n_lB{\chi}_\alpha gg_*G_\alpha (g^\prime )G_\alpha (g^\prime _*)dvdv_*dn dxds\\
= &\int _0^T\int (v_1-v_{*1})\sum^3_{l=2}(v_l-v_{*l})n_1n_lB{\chi}_\alpha gg_*G_\alpha (g^\prime )G_\alpha (g^\prime _*)dvdv_*dn dxds\\
\leq &\frac{1}{\beta ^2}\int _0^T\int (v_1-v_{*1})^2B{\chi}_\alpha gg_*G_\alpha (g^\prime )G_\alpha (g^\prime _*)dvdv_*dn dxds\\
&+\frac{{\beta ^2}}{4}\int _0^T\int \sum^3_{l=2}(v_l-v_{*l})^ 2n^2_1n_l^2B{\chi}_\alpha gg_*G_\alpha (g^\prime )G_\alpha (g^\prime _*)dvdv_*dn dxds\\
\leq &\frac{2c'}{{\beta ^2}}{(1+T)}(g_\infty +1)^2+\frac{{\beta ^2}}{4}\int _0^T\int n_1^2 \sum^3_{l=2}(v_l-v_{*l})^2n_l^2B{\chi}_\alpha gg_*G_\alpha (g^\prime )G_\alpha (g^\prime _*)dvdv_*dn dxds,
\end{aligned}\]
for any $\beta >0$. It follows that
\begin{eqnarray*}
\int_0^T\int n_1^2[(v-v_*)\cdot n]^2B{\chi}_\alpha gg_*G_\alpha (g^\prime )G_\alpha (g^\prime _*)dvdv_*dn dxds\leq c'_0(g_\infty +1)^2(1+T),
\end{eqnarray*}
with $c'_0$ only depending on $\int f_0(x,v)dxdv$ and $\int \lvert v\rvert ^2f_0(x,v)dxdv$. This completes the proof of the lemma. \cqfd
Denote by $M_\alpha $ the mass density
\begin{eqnarray*}
M_\alpha (t)= \int \sup _{(s,x)\in [ 0,t] \times [ 0,1]}g^\sharp _\alpha (s,x,v)dv.
\end{eqnarray*}
Lemmas \ref{integral-dxdv-local} to \ref{first-T-dependent-on-alpha} are devoted to the local in time uniform control with respect to $\alpha $ of the mass densities $(M_\alpha )$.
%
%
\begin{lemma}\label{integral-dxdv-local}
\hspace*{0.2in}\\
For any $\epsilon>0$, there exists a constant $c'_1$ only depending on
$\int f_0(x,v)dxdv$ and $\int |v|^2f_0(x,v)dxdv$, such that on any interval of time $[0,T]$ where the solution $g_\alpha $ to (\ref{eq-g-alpha}) is bounded by $g_\infty $,
\begin{equation}\label{dxdv-local}
\int \sup_{s\in [ 0,t] }g_\alpha ^\sharp (s,x,v)dxdv\leq c'_1(g_\infty +1)^2\Big( (1+\frac{1}{\epsilon ^2})(1+t)+\epsilon tM_\alpha (t)\Big) ,\quad t\in [0,T],\quad \alpha \in ] 0,1[.
\end{equation}
\end{lemma}
\hspace*{1.in}\\
\underline{Proof of Lemma \ref{integral-dxdv-local}.} \\
Denote $g_\alpha $ by $g$ for simplicity. Given its nonnegative values it satisfies
\begin{align}
&g^{\sharp }(t,x,v)= f_{0,\alpha }(x,v)+\int_0^t\tilde{Q}_\alpha(g)(s,x+sv_1,v)ds\nonumber \\
&\leq f_{0,\alpha }(x,v)+\int_0^t\tilde{Q}^{+}_\alpha (g)(s,x+sv_1,v)ds= f_{0,\alpha }(x,v)\nonumber \\
&+\int_0^t\int Bg(s,x+sv_1,v^\prime )g(s,x+sv_1,v^\prime _*)
G_\alpha (g)(s,x+sv_1,v)G_\alpha (g)(s,x+sv_1,v_*)dv_*dnds.\label{control-g-alpha-by-gain}
 \end{align}
For any $(v,v_*)\in \R ^3\times \R ^3$, let $\mathcal{N}_\epsilon$ be the set of $n\in \mathcal{S} ^2$ with $\max \{ n_1,n_{\perp 1}\} <\epsilon$, where $n_\perp $ is the unit vector in the direction $v-v^\prime _*$ (orthogonal to $n$) in the plane defined by $v-v_*$ and $n$, and $n_1$ is the component of $n$ along the $x$-axis.
Let $\mathcal{N}^c _\epsilon$ be the complement of $\mathcal{N}_\epsilon$ in $\mathcal{S}^2$.  Denote by
\begin{eqnarray*}
\mathcal{I}_\epsilon (t)= \int_0^t\int \int _{\mathcal{N}_\epsilon} B{\chi}_\alpha
g(r,x+rv_1,v^\prime )g(r,x+rv_1,v^\prime _*)G_\alpha (g)^\sharp(r,x,v)G_\alpha (g)(r,x+rv_1,v_*)dndvdv_*dxdr.
\end{eqnarray*}
(\ref{ineq-bony}) also holds with $n_1$ replaced by $n_{\perp 1}$. Integrating (\ref{control-g-alpha-by-gain}) with respect to $(x,v)$ and using Lemma \ref{bony} leads to
\begin{align}\label{lemma4.1-1}
&\int \sup_{s\in [0, t]}g^\sharp (s,x,v)dxdv
\leq  \int f_0(x,v)dxdv+\mathcal{I}_\epsilon (t)\nonumber \\
&+\int_0^t\int \int _{\mathcal{N}^c _\epsilon} B{\chi}_\alpha
g(r,x+rv_1,v^\prime )g(r,x+rv_1,v^\prime _*)G_\alpha (g)^\sharp (r,x,v)G_\alpha (g)(r,x+rv_1,v_*)dvdv_*dn dxdr\nonumber \\
&\leq  \int f_0(x,v)dxdv
+\mathcal{I}_\epsilon (t)+\frac{1}{(\gamma \gamma ^\prime \epsilon )^2}\int_0^t\int (n_1^2+n_{\perp 1}^2)[(v-v_*)\cdot n]^2B{\chi}_\alpha gg_*G_\alpha (g^\prime )G_\alpha (g^\prime _*)dvdv_*dn dxdr\nonumber \\
&\leq  \int f_0(x,v)dxdv+\mathcal{I}_\epsilon (t)+\frac{2c_0^\prime }{(\gamma \gamma ^\prime \epsilon )^2}(g_\infty +1)^2(1+t).
\end{align}
Moreover,
\begin{eqnarray*}
\mathcal{I}_\epsilon (t)\leq c\epsilon \hspace*{0.03in}tM_\alpha (t)(g_\infty +1)^2\int f_0(x,v)dxdv.
\end{eqnarray*}
The lemma follows.\cqfd
%
%
\begin{lemma}\label{small-set-local}
\hspace*{0.2in}\\
For any $\delta \in ] 0,1[ $, there is $c_2^\prime $  only depending on $\int f_0(x,v)dxdv$ and $\int |v|^2f_0(x,v)dxdv$, {such} that on any interval of time $[0,T]$ where the solution $g_\alpha $ to (\ref{eq-g-alpha}) is bounded by $g_\infty $,
\begin{align}\label{control-small-set-local}
&\sup _{x_0\in[0,1] }\int_{|x-x_0|<\delta } \hspace*{0.03in}{\sup_{s\in [ 0,t] }}g_\alpha ^\sharp (s,x,v)dxdv\leq c_2^\prime (g_\infty +1)^2\Big( \delta ^{\frac{2}{5}}+t^{\frac{8}{11}}(1+t)^{\frac{3}{11}}\big(1+M_\alpha (t)\big) \Big) ,\nonumber \\
&\hspace*{4.2in} t\in [0,T],\quad \alpha \in ] 0,1[ .
\end{align}
\end{lemma}
\underline{Proof of Lemma \ref{small-set-local}.} \\
Denote $g_\alpha $ by $g$ for simplicity. For $s\in [0,t] $ it holds,
\[ \begin{aligned}
 g^\sharp (s,x,v)&=g^{\sharp}(t,x,v)-\int_{s}^t\tilde{Q}_\alpha (g)^\sharp (r,x,v)dr\leq g^{\sharp}(t,x,v)+\int_{s}^t(\tilde{Q}_\alpha ^-(g))^\sharp (r,x,v)dr,
\end{aligned}\]
where $\tilde{Q}_\alpha ^-$ is defined in (\ref{Qtildealpha-}). And so
\begin{align}
&\sup_{s\in [ 0,t] }g^\sharp (s,x,v)\leq  g^{\sharp}(t,x,v)\nonumber \\
&+\int_0^t\int B{\chi}_\alpha g^\sharp (r,x,v)g(r,x+rv_1 ,v_*)G_\alpha (g)(r,x+rv_1,v^\prime )G_\alpha (g)(r,x+rv_1,v^\prime _*)dv_*dn dr.\label{Qminus-local}
\end{align}
Denote by
\begin{align*}
\mathcal{J}_\epsilon (t)= \sup _{x_0\in [ 0,1] }\int_0^{t}\int_{|x-x_0|<\delta }\int\int_{\mathcal{N}_\epsilon} B{\chi}_\alpha
&g^\sharp (r,x,v)g(r,x+rv_1 ,v_*)\\
&G_\alpha (g)(r,x+rv_1,v^\prime )G_\alpha (g)(r,x+rv_1,v^\prime _*)dvdv_*dn dxdr.
\end{align*}
Integrating (\ref{Qminus-local}) with respect to $(x,v)${, using Lemma \ref{bony} and the $g_\infty $ bound of $g$}, gives for any $x_0\in [0,1]$, $\lambda >0$ and $\Lambda >0$ that
\begin{align}\label{lemma4.2-1}
&\int_{|x-x_0|<\delta } \sup_{s\in [ 0,t] }g^\sharp (s,x,v)dxdv\leq  \int_{|x-x_0|<\delta } g^{\sharp}(t,x,v)dxdv+\mathcal{J}_\epsilon (t)\nonumber \\
&+\frac{1}{(\lambda \gamma ^{\prime }\epsilon )^2}\int_0^t\int_{|v-v_*|\geq\lambda} (n_1^2+n_{\perp 1}^2)[(v-v_*)\cdot n]^2B{\chi}_\alpha gg_*G_\alpha (g^\prime )G_\alpha (g^\prime _*)dvdv_*dn dxds\nonumber \\
&+c(g_\infty +1)^2\int_{0}^{t}\int _{|v-v_*|<\lambda} B{\chi}_\alpha g^{\sharp}(s,x,v)g(s,x+sv_1,v_*)dvdv_*dn dxds\nonumber \\
\leq  &\int_{|x-x_0|<\delta } g^{\sharp}(t,x,v)dxdv+\mathcal{J}_\epsilon (t)
+\frac{c_0^\prime (1+t)(g_\infty +1)^2}{(\lambda \gamma ^\prime \epsilon )^2}+ct \lambda^3(g_\infty +1)^2\int f_0(x,v)dxdv\nonumber \\
\leq &\frac{1}{\Lambda ^2}\int v^2f_0dxdv + c\delta \Lambda^3+\mathcal{J}_\epsilon(t)+\frac{c_0^\prime (1+t)(g_\infty +1)^2}{(\lambda \gamma ^\prime \epsilon )^2}+ct\lambda^3(g_\infty +1)^2\int f_0 (x,v)dxdv\nonumber \\
\leq &c(g_\infty +1)^2\big( \delta ^{\frac{2}{5}}+t^{\frac{2}{5}}\epsilon ^{-\frac{6}{5}}(1+t)^{\frac{3}{5}}\big)  +\mathcal{J}_\epsilon(t),
\end{align}
for an appropriate choice of $(\Lambda ,\lambda )$. Moreover, $\mathcal{J}_\epsilon (t)\leq c(g_\infty +1)^2\epsilon tM_\alpha (t)$.
Taking $\epsilon = \tilde{c}\big( \frac{1+t}{t}\big) ^{\frac{3}{11}}M_\alpha ^{-\frac{5}{11}}$ with $\tilde{c}$ suitably chosen, leads to the statement of the lemma.\cqfd
%
%
\begin{lemma}\label{mass-tails-depending-on-alpha}
\hspace*{0.02in}\\
The solution $g_\alpha $ to (\ref{eq-g-alpha}) satisfies
\begin{eqnarray*}
\int _0^1\int_{|v|>\lambda} |v|\sup_{s\in [ 0,t] }g_\alpha ^\sharp (s,x,v)dvdx\leq\frac{ct}{\lambda \alpha ^2}M_\alpha (t),\quad t\in [0,T],
\end{eqnarray*}
where $c$ only depends on $\int |v|^2f_0(x,v)dxdv$.
\end{lemma}
\underline{Proof of Lemma \ref{mass-tails-depending-on-alpha}.} \\
Denote $g_\alpha $ by $g$ for simplicity. Multiply (\ref{control-g-alpha-by-gain}) by $\lvert v\rvert $ and integrate over $[0,1] \times \{ v\in \R ^3; |v|>\lambda \}$. It results,
\begin{align*}
&\int _0^1\int_{|v|>\lambda}|v|\sup_{s\in [ 0,t] }g_\alpha ^\sharp (s,x,v)dvdx
\leq  \int \int_{|v|>\lambda}|v|f_{0,\alpha }(x,v)dvdx+\int_0^t\int_{|v|>\lambda} B{\chi}_\alpha \\
&|v| g(s,x+sv_1,v^\prime )g(s,x+sv_1,v^\prime _*)G_\alpha (g)(s,x+sv_1,v)G_\alpha (g)(s,x+sv_1,v_*)dvdv_*dndxds.
\end{align*}
Here in the last integral, either $|v^\prime |$ or $|v^\prime _*|$ is the largest and larger than $\frac{\lambda}{\sqrt 2}$. The two cases are symmetric, and we discuss the case $|v^\prime |\geq|v^\prime _*|$. After a translation in $x$, and using that $G_\alpha (g_\alpha )$ is bounded by $\frac{1}{\alpha }$, the integrand of the r.h.s of the former inequality is estimated {from above} by
\begin{eqnarray*}
\frac{c}{\alpha ^2} |v^\prime |g^{\#} (s,x,v^\prime )\sup_{(s,x)\in [ 0,t] \times [0,1]}g^{\#}(s,x,v^\prime _*).
\end{eqnarray*}
The change of variables {$(v,v_*,n)\rightarrow (v^\prime ,v^\prime _*,-n)$} and the integration over
 \begin{eqnarray*}
(s,x,v,v_*,n)\in [ 0,t] \times [ 0,1] \times \{ v\in \R ^3; |v| >\frac{\lambda}{\sqrt 2}\} \times \R ^3 \times \mathcal{S}^2,
\end{eqnarray*}
give the bound
\begin{align*}
&\frac{c}{\lambda \alpha ^2}\Big( \int_0^{t}\int |v|^2g^{\#}(s,x,v)dxdvds\Big) \Big( \int \sup_{(s,x)\in [ 0,t] \times [0,1]} g^{\#}(sx,v_*)dv_*\Big) \\
&=\frac{ctM_\alpha (t)}{\lambda \alpha ^2}\int |v|^2f_0(x,v)dxdv.
\end{align*}
The lemma follows.                        \cqfd
%
%
\begin{lemma}\label{first-T-dependent-on-alpha}
\hspace*{0.1in}\\
Let $c_0= \int \sup _{x\in [ 0,1] }f_0(x,v)dv$. There is $T_\alpha >0$ such that the solution $g_\alpha $ to (\ref{eq-g-alpha}) satisfies
\begin{eqnarray*}
\int \sup_{(t,x)\in [0,T_\alpha ] \times [ 0,1] }g^\sharp _\alpha (t,x,v)dv\leq 2c_0.
\end{eqnarray*}
\end{lemma}
\underline{Proof of Lemma \ref{first-T-dependent-on-alpha}.} \\
Denote $g_\alpha $ by $g$. Denote by $E(x)$ the integer part of $x\in\R$, $E(x)\leq x< E(x)+1$. By (\ref{control-g-alpha-by-gain}),
\begin{align}\label{in1-local}
&\sup_{s\in [ 0,t] }g^\sharp (s,x,v)\leq  f_{0,\alpha }(x,v)+\frac{1}{\alpha ^2}(A_1+A_2+A_3+A_4),
\end{align}
where, for $\epsilon >0$, $\delta  >0$ and $\lambda $ that will be fixed later,
\begin{align*}
&A_1= \int_0^t\int _{|n_1| \geq \epsilon ,\hspace*{0.02in}t|v_1-v^\prime _1|>\delta  }B{\chi}_\alpha \sup _{\tau \in [ 0,t]} g^{\#}(\tau ,x+s(v_1-v^\prime _1),v^\prime ) \sup _{\tau \in [ 0,t]}g^{\#}(\tau ,x+s(v_1-v^\prime _{*1}),v^\prime _*)dv_*dn ds,\\
&A_2= \int_0^t\int _{|n_1| \geq \epsilon ,\hspace*{0.02in}t|v_1-v^\prime _1|<\delta ,\hspace*{0.02in} |v^\prime |<\lambda }B{\chi}_\alpha \sup _{\tau \in [ 0,t]} g^{\#}(\tau ,x+s(v_1-v^\prime _1),v^\prime ) \times \\
& \hspace*{2.3in}\times \sup _{\tau \in [ 0,t]}g^{\#}(\tau ,x+s(v_1-v^\prime _{*1}),v^\prime _*)dv_*dn ds,\\
&A_3= \int_0^t\int _{|n_1| \geq \epsilon ,\hspace*{0.02in}t|v_1-v^\prime _1|<\delta ,\hspace*{0.02in} |v^\prime |>\lambda }B{\chi}_\alpha \sup _{\tau \in [ 0,t]} g^{\#}(\tau ,x+s(v_1-v^\prime _1),v^\prime ) \times \\
& \hspace*{2.3in}\times \sup _{\tau \in [ 0,t]}g^{\#}(\tau ,x+s(v_1-v^\prime _{*1}),v^\prime _*)dv_*dn ds,\\
&A_4= \int_0^t\int _{|n_1|<\epsilon }B{\chi}_\alpha \sup _{\tau \in [ 0,t]} g^{\#}(\tau ,x+s(v_1-v^\prime _1),v^\prime ) \sup _{\tau \in [ 0,t]}g^{\#}(\tau ,x+s(v_1-v^\prime _{*1}),v^\prime _*)dv_*dn ds.
\end{align*}
In $A_1$, $A_2$ and $A_3$, bound the factor $\sup _{\tau \in [ 0,t] }g^\sharp (\tau ,x+s(v_1-v^\prime _{*1}),v^\prime _*)$ by its supremum over $x\in [ 0,1] $, and make the change of variables
\begin{eqnarray*}
 s\rightarrow y= x+s(v_1-v^\prime _1),
\end{eqnarray*}
with Jacobian
\begin{eqnarray*}
\frac{Ds}{Dy}= \frac{1}{|v_1-v^\prime _1|}= \frac{1}{|v-v_*|\hspace*{0.03in}|(n,\frac{v-v_*}{|v-v_*|})|\hspace*{0.03in}|n_1|} \leq \frac{1}{\epsilon \gamma \gamma ^\prime }\hspace*{0.02in}.
\end{eqnarray*}
Consequently,
\begin{align*}
&\sup _{x\in [0,1] }A_1(t,x,v)\\
&\leq \sup _{x\in [0,1] }\int _{t|v_1-v^\prime _1|>\delta }\frac{B{\chi}_\alpha }{|v_1-v^\prime _1|}\Big( \int _{y\in (x,x+t(v_1-v^\prime _1))}\sup _{\tau \in [ 0,t]} g^{\#}(\tau ,y,v^\prime )dy\Big)  \sup _{(\tau ,X)\in [ 0,t]\times [ 0,1] }g^{\#}(\tau ,X,v^\prime _*)dv_*dn \\
&\leq \int _{t|v_1-v^\prime _1|>\delta }\frac{B{\chi}_\alpha }{|v_1-v^\prime _1|} | E(t(v_1-v^\prime _1)+1)|\Big( \int _0^1\sup _{\tau \in [ 0,t]} g^{\#}(\tau ,y,v^\prime )dy\Big)  \sup _{(\tau ,X)\in [ 0,t]\times [ 0,1] }g^{\#}(\tau ,X,v^\prime _*)dv_*dn.
\end{align*}
Performing the change of variables $(v,v_*,n )\rightarrow (v^\prime ,v^\prime _*,-n )$,
\begin{align}
&\int \sup _{x\in [ 0,1] }A_1(t,x,v)dv\nonumber \\
&\leq \int _{t|v_1-v^\prime _1|>\delta }\frac{B{\chi}_\alpha }{|v_1-v^\prime _1|} | E(t(v^\prime _1-v_1)+1)|\Big( \int _0^1\sup _{\tau \in [ 0,t]} g^{\#}(\tau ,y,v)dy\Big)  \sup _{(\tau ,X)\in [ 0,t]\times [ 0,1] }g^{\#}(\tau ,X,v_*)dvdv_*dn \nonumber \\
&\leq t(1+\frac{1}{\delta })\int B{\chi}_\alpha \Big( \int _0^1\sup _{\tau \in [ 0,t]} g^{\#}(\tau ,y,v)dy\Big)  \sup _{(\tau ,X)\in [ 0,t]\times [ 0,1] }g^{\#}(\tau ,X,v_*)dvdv_*dn\nonumber \\
&\leq 4\pi B_0\hspace*{0.02in} t(1+\frac{1}{\delta})\Big( \int \sup _{\tau \in [ 0,t]} g^{\#}(\tau ,y,v)dydv\Big) M_\alpha (t).\nonumber
\end{align}
Apply Lemma \ref{integral-dxdv-local} with $g_\infty = \frac{1}{\alpha }$, so that
\begin{equation}\label{bdd-A1-local}
\int \sup _{x\in [ 0,1] }A_1(t,x,v)dv\leq \frac{4\pi B_0c_1^\prime t}{\alpha ^2} (1+\frac{1}{\delta })\Big( (1+\frac{1}{\epsilon ^2})(1+t)+\epsilon tM_\alpha (t)\Big) M_\alpha (t) .
\end{equation}
%
%
Moreover,
\begin{align}
c\hspace*{0.01in}\epsilon \int \sup _{x\in [0,1] }A_2(t,x,v)dv&\leq \frac{\delta }{\alpha } \int \int _{|v^\prime |<\lambda }B\chi _\alpha \sup _{(\tau ,X)\in [ 0,t] \times [ 0,1] } g^{\#}(\tau ,X,v^\prime _*)dvdv_*dn\nonumber \\
&= \frac{\delta }{\alpha } \int _{|v|<\lambda }\int B\chi _\alpha \sup _{(\tau ,X)\in [ 0,t] \times [ 0,1] } g^{\#}(\tau ,X,v_*)dvdv_*dn\nonumber \\
& \hspace*{1.in}\text{by the change of variables }(v,v_*,n)\rightarrow (v^\prime ,v^\prime _*,-n)\nonumber \\
&\leq c\frac{\delta }{\alpha }\lambda ^3M_\alpha (t), \label{A2-local}
\end{align}
and
\begin{align}
c\hspace*{0.01in}\epsilon \int \sup _{x\in [0,1] }A_3(t,x,v)dv&\leq \int _{|v^\prime |>\lambda }B\chi _\alpha \Big( \int _0^1\sup _{\tau \in [ 0,t] } g^{\#}(\tau ,y,v^\prime )dy\Big) \sup _{(\tau ,X)\in [ 0,t] \times [ 0,1] } g^{\#}(\tau ,X,v^\prime _*)dvdv_*dn\nonumber \\
&\leq c\Big( \int _0^1\int _{|v|>\lambda }\sup _{\tau \in [ 0,t] } g^{\#}(\tau ,y,v)dvdy\Big) \int \sup _{(\tau ,X)\in [ 0,t] \times [ 0,1] } g^{\#}(\tau ,X,v_*)dv_*\nonumber \\
&\hspace*{1.in}\text{by the change of variables }(v,v_*,n)\rightarrow (v^\prime ,v^\prime _*,-n)\nonumber \\
&\leq \frac{ct}{\lambda \alpha ^2}M_\alpha ^2(t)\quad \text{by Lemma \ref{mass-tails-depending-on-alpha}}. \label{A3-local}
\end{align}
Finally, with the change of variables $(v,v_*,n)\rightarrow (v^\prime ,v^\prime _*,-n)$,
\begin{align}
\int \sup _{x\in [0,1] }A_4(t,x,v)dv&\leq B_0t\big( \int _{|n_1| <\epsilon }dn\big) \Big( \int \sup _{(\tau ,x)\in [ 0,t] \times [ 0,1] } f^{\#}(\tau ,x,v)dv\Big) ^2\nonumber \\
& \leq 2\pi B_0\epsilon \hspace*{0.02in}t M_\alpha ^2(t).\label{A4-local}
\end{align}
It follows from (\ref{in1-local}), (\ref{bdd-A1-local}), (\ref{A2-local}), (\ref{A3-local}) and (\ref{A4-local}) that
\hspace*{0.1in}\\
\begin{equation}\label{trinome-M}
a(t)M^2_\alpha (t)-b(t)M_\alpha (t)+c_{0,\alpha }\geq 0,\quad t\leq 1,
\end{equation}
where $c_{0,\alpha }= \int \sup _{x\in [0,1]}f_{0,\alpha }(x,v)dv$, and for some positive constants $(c^\prime _l)_{2\leq l\leq 4}$ independent on $\epsilon $, $\delta $, $\lambda $ and $\alpha $,
\begin{align*}
&a(t)= \frac{c^\prime _2\hspace*{0.01in} t}{\alpha ^2}\Big( \epsilon +\epsilon ^{-1}\lambda ^{-1}\alpha ^{-2}+\epsilon t(1+\delta ^{-1})\alpha ^{-2}\Big) ,\\
&b(t)= 1-c^\prime _3t(1+\delta ^{-1})(1+\epsilon ^{-2})\alpha ^{-4}-c^\prime _4\delta \epsilon ^{-1}\alpha ^{-3}\lambda ^3.
\end{align*}
Choose $\epsilon = \frac{\alpha ^3}{16c_4^\prime }$, $\delta = \epsilon ^8$ and $\lambda = \epsilon ^{-2}$. For $T$ small enough depending on $\alpha $, it holds that
\begin{equation}\label{sufficient-local}
b(t)\in \hspace*{0.03in}] \frac{3}{4}, 1[ \quad \text{and   }c_{0,\alpha }a(t)<\frac{1}{8},\quad t\in [0,T] ,
\end{equation}
which is sufficient  for the polynomial in (\ref{trinome-M}) to have two nonnegative roots and take a negative value at $2c_{0,\alpha }$. Recalling that  $M_\alpha (0)= c_{0,\alpha }$ and $M_\alpha $ is a continuous function, it follows that
\begin{equation}\label{time-of-control-of-M}
M_\alpha (t)\leq 2c_{0,\alpha }\leq 2c_0, \quad t\in [0,T] .
\end{equation}
\cqfd
%
%
%
\begin{lemma}\label{second-T-dependent-on-alpha}
\hspace{.1cm}\\
Given $f_0\leq2^L$ satisfying (\ref{hyp-f0-1})-(\ref{hyp-f0-2}), there is for each $\alpha \in ]0,2^{-L-1}[$ a time $\bar{T}_\alpha >0$ so that the solution $g_\alpha$ to (\ref{eq-g-alpha}) is bounded by $2^{L+1}$ on $[0,\bar{T}_\alpha ]$.
\end{lemma}
\underline{Proof of Lemma \ref{second-T-dependent-on-alpha}.}\\
Denote $g_\alpha $ by $g$ for simplicity. By (\ref{control-g-alpha-by-gain}), the $\frac{1}{\alpha }$ bound of $g$ and $G_\alpha (g)$,
\begin{align*}
&\sup_{s\in [ 0,t] }g^{\sharp} (s,x,v)
\leq  2^L+\frac{B_0}{\alpha ^3}\int _0^t\int g(s,x+sv_1,v^\prime )dv_*dnds.
\end{align*}
With the angular cut-off (2.2), $v_*  \rightarrow v^\prime $ is a change of variables. Using it and (\ref{first-T-dependent-on-alpha}) leads for some constant $c>0$ to
\begin{align*}
\sup_{(s,x)\in [0,t] \times [0,1]}g^{\sharp} (s,x,v)&\leq 2^L+ \frac{2B_0cc_0}{\alpha ^3}\hspace*{0.02in}t\\
&\leq 2^{L+1}\hspace*{1.2in}\text{for }t\leq \min \{T_\alpha , \frac{2^{L-1}\alpha ^3}{B_0cc_0}\} \, .
\end{align*}
The lemma follows.\cqfd
%
%
\begin{lemma}\label{T-L-independent-on-alpha}
\hspace{.1cm}\\
Given $f_0\leq2^L$ satisfying (\ref{hyp-f0-1})-(\ref{hyp-f0-2}), there is $T_L>0$ such that for all $\alpha \in ] 0,2^{-L-1}[ $, the solution $g_\alpha$ to (\ref{eq-g-alpha}) is bounded by $2^{L+1}$ on $[0,T_L]$ and
\begin{equation}\label{T-independent-on-alpha}
\int \sup _{(t,x)\in [0,T_L] \times [0,1] }g_\alpha (t,x,v)dv\leq 2c_0.
\end{equation}
\end{lemma}
\underline{Proof of Lemma \ref{T-L-independent-on-alpha}.}\\
Given $\alpha\leq 2^{-L-1}$, it follows from Lemmas \ref{first-T-dependent-on-alpha} and \ref{second-T-dependent-on-alpha} that the maximum time $T^\prime _\alpha $ for which $g_\alpha \leq 2^{L+1}$ on $[0,T^\prime _\alpha ]$ and $\int \sup _{(t,x)\in [0,T_\alpha ^\prime ] \times [0,1] }g_\alpha (t,x,v)dv\leq 2c_0$ is positive. Following the lines of the proof of Lemma \ref{first-T-dependent-on-alpha} with all bounds on $g_\alpha $ and $G_\alpha $ now depending on $L$ (from the $2^{L+1}$ bound on $g_\alpha $), one gets (\ref{trinome-M}) with $a(t)$, $b(t)$ and $c(t)$ depending on $L$. Consequently the same argument leading to the existence of a positive time $T_L^\prime $ so that $M_\alpha (\min \{ T_\alpha ^\prime ,T_L^\prime \} )\leq \frac{3}{2}c_0$ holds. Moreover, by (\ref{control-g-alpha-by-gain}), the change of variables $v_*  \rightarrow v^\prime $, and the $2c_0$ bound for $M_\alpha (\min \{ T_\alpha ^\prime ,T_L^\prime \} )$,
\begin{align*}
\sup_{(s,x)\in [0,t] \times [0,1] }g_{\alpha }^{\sharp} (s,x,v)&\leq f_{0,\alpha }(x,v)+ cB_02^{3L}t\int \sup_{(s,x)\in [0,t] \times [0,1] }g_{\alpha }(s,x,v')dv'\\
&\leq 2^L+cc_0B_02^{3L+1}t\\
&\leq  3(2^{L-1})\hspace*{0.8in} , t\in [ 0, \min \{ T_\alpha ^\prime ,T^\prime _L,\frac{1}{cc_0B_02^{2L+2}} \} ] \, .
\end{align*}
For all $\alpha\leq 2^{-L-1}$, it holds that $T^\prime _\alpha \geq \min \{ T_L^\prime ,\frac{1}{cc_0B_02^{2L+2}} \} $, else $T^\prime _\alpha $ would not be the maximum time such that $g_\alpha (t)\leq 2^{L+1}$ on $[ 0,T^\prime _\alpha ]$ and $\int \sup _{(t,x)\in [0,T_\alpha ^\prime ] \times [0,1] }g_\alpha (t,x,v)dv\leq 2c_0$. Denote by $T_L= \min \{ T_L^\prime ,\frac{1}{cc_0B_02^{2L+2}} \}$. The lemma follows since $T_L$ does not depend on $\alpha $.\cqfd\\
The control of the mass density of $g_\alpha $ over large velocities is performed in the following lemma.\\
%
%
\begin{lemma}\label{large-velocity-2}
\hspace*{0.02in}\\
The solution $g_\alpha $ to (\ref{eq-g-alpha}) satisfies
\begin{equation}\label{g-alpha-large-velocities}
\int_{|v|>\lambda} \sup_{(s,x)\in [ 0,T_L] \times [0,1]}g_\alpha ^\sharp (s,x,v)dv\leq\frac{cT_L}{\sqrt{\lambda}},
\end{equation}
for some constant $c>0$.
\end{lemma}
\underline{Proof of Lemma \ref{large-velocity-2}.}\\
Denote $g_\alpha $ by $g$ for simplicity. Following the lines of the proof of Lemma \ref{mass-tails-depending-on-alpha} with the bounds on $g_\alpha $ and $G_\alpha (g_\alpha )$ no more depending on $\alpha $ but on $L$, it holds that
\begin{equation}\label{large-tails-L}
\int _0^1\int_{|v|>\lambda} |v|\sup_{t\in [ 0,T_L] }g_\alpha ^\sharp (t,x,v)dvdx\leq\frac{c_L}{\lambda}M_\alpha (T_L),
\end{equation}
where $c_L$ only depends on $L$ and $\int |v|^2f_0(x,v)dxdv$.
 By (\ref{control-g-alpha-by-gain}) with the $2^{L+1}+1$ of $G_\alpha (g)$ used,
\begin{eqnarray}
 \int _{|v|>\lambda }\sup_{(s,x)\in [ 0,T_L] \times [ 0,1] }g^\sharp (s,x,v)dv
\leq  \int _{|v|>\lambda }\sup_{x\in [ 0,1] }f_0(x,v)dv+(2^{L+1}+1)^2C,
\end{eqnarray}
where
\begin{eqnarray*}
C= \int _{|v|>\lambda }\sup _{x\in [ 0,1] }\int_0^{T_L}\int B
 g^{\#}(s,x+s(v_1-v^\prime _1),v^\prime )g^{\#}(s,x+s(v_1-v^\prime _{*1}),v^\prime _*)dvdv_*dnds.
\end{eqnarray*}
For $v',v'_*$ outside of the angular cutoff {(2.2)}, let $n$ be the unit vector in the direction $v-v'$ and $n_\perp $ its orthogonal unit vector in the direction $v-v^\prime _*$. Split $C$ into $C= \sum _{0\leq i\leq 2}C_i$, where
\begin{eqnarray*}
C_0= \int _{|v|>\lambda }\sup _{x\in [ 0,1] }\Big( \int_0^{T_L}\int _{|n_1|<\epsilon \text{  or  } |n_{\perp 1}|<\epsilon }B
 g^{\#}(s,x+s(v_1-v^\prime _1),v^\prime )g^{\#}(s,x+s(v_1-v^\prime _{*1}),v^\prime _*)dv_*dnds\Big) dv,
\end{eqnarray*}
and $C_1$ (resp. $C_2$) refers to integration {with respect to $(v_*,n)$ on}
\begin{align*}
&\{ (v_*,n); \quad |n_1|\geq \epsilon, \quad |n_{\perp 1}|\geq \epsilon, \quad   |v^\prime | \geq |v^\prime _*|\} ,\\
\big( \text{resp.    } &\{ (v_*,n); \quad |n_1|\geq \epsilon, \quad |n_{\perp 1}|\geq \epsilon, \quad   |v^\prime | \leq |v^\prime _*|\} \big) .
\end{align*}
By (\ref{T-independent-on-alpha}) and the change of variables $(v,v_*,n)\rightarrow (v^\prime ,v^\prime _*,-n)$,
\begin{equation}\label{C0}
C_0\leq c\epsilon T_L,
\end{equation}
for some constant $c>0$. Moreover, by the change of variables $s\rightarrow y= x+s(v_1-v^\prime _1)$,
\begin{align*}
C_1&\leq \int _{|v|\geq \lambda }\sup _{x\in [ 0,1] }\int _{|v^\prime |>|v^\prime _*|}B(\int _0^{T_L}\sup _{\tau \in [0,T_L]}g^\sharp (\tau ,x+s(v_1-v_1^\prime ),v^\prime )ds)\sup _{(\tau ,X)\in [0,T_L] \times [0,1]}g^\sharp (\tau ,X,v^\prime _*)dvdv_*dn\\
&=  \int _{|v|\geq \lambda }\sup _{x\in [ 0,1] }\int _{|v^\prime |>|v^\prime _*|}\frac{B}{|v_1-v^\prime _1|}(\int _{y\in (x,x+T_L(v_1-v^\prime _1)}\sup _{\tau \in [0,t]}g^\sharp (\tau ,y,v^\prime )dy)\sup _{(\tau ,X)\in [0,t] \times [0,1]}g^\sharp (\tau ,X,v^\prime _*)dvdv_*dn\\
&\leq \int _{|v|\geq \lambda , |v^\prime |>|v^\prime _*|}B\frac{E(T_L|v_1-v^\prime _1|)+1}{|v_1-v^\prime _1|}(\int _0^1\sup _{\tau \in [0,t]}g^\sharp (\tau ,y,v^\prime )dy)\sup _{(\tau ,X)\in [0,t] \times [0,1]}g^\sharp (\tau ,X,v^\prime _*)dvdv_*dn\\
&\leq c(T_L+\frac{1}{\epsilon \gamma \gamma ^\prime })\int _{|v|\geq \frac{\lambda }{\sqrt{2}}}\int _0^1\sup _{\tau \in [0,t]}g^\sharp (\tau ,y,v)dydv\\
&\leq \frac{c}{\lambda }(T_L+\frac{1}{\epsilon }),\quad \text{by  }(\ref{large-tails-L}).
\end{align*}
The term $C_2$ can be controlled similarly to $C_1$ with the change of variables $s\rightarrow y= x+s(v_1-v^\prime _{*1})$. And so,
\begin{align*}
&C\leq c\big( \epsilon T_L+\frac{1}{\lambda }(T_L+\frac{1}{\epsilon})\big) .
\end{align*}
Choosing $\epsilon = \frac{1}{\sqrt{\lambda }}$ leads to
\begin{align*}
&C\leq \frac{cT_L}{\sqrt{\lambda }}.
\end{align*}
\hspace*{0.1in}\\
The lemma follows. \cqfd
%
%
%
%
\section{Proof of Theorem 1.1.}
Let us first prove that $(g_\alpha )$ is a Cauchy sequence in $C([0,T_L]; L^1([0,1] \times \R ^3))$ with $T_L$ defined in Lemma \ref{T-L-independent-on-alpha}.
For any $(\alpha _1,\alpha _2)\in ]0,2^{-L-1}[ ^2$, the function $g= g_{\alpha_1}-g_{\alpha_2}$ satisfies the equation
\[ \begin{aligned}
\partial _tg+v_1\partial _xg&= \int  B(g_{\alpha_1}^{\prime} g_{\alpha_1*}^{\prime } -g_{\alpha_2}^{\prime} g_{\alpha_2 *}^{\prime } )G_{\alpha_1}(g_{\alpha_1})G_{\alpha_1}(g_{\alpha_1 *})dv_*dn \\
&-\int  B(g_{\alpha_1}g_{\alpha_1 *} -g_{\alpha _2}g_{\alpha_2 *})G_{\alpha_1}(g_{\alpha_1}^{\prime } )G_{\alpha_1}(g_{\alpha_1 *}^{\prime } )dv_*dn \\
&+\int Bg_{\alpha_2}^{\prime } g_{\alpha_2 *}^{\prime } \Big( G_{\alpha_1}(g_{\alpha_1 *})\big( G_{\alpha_1}(g_{\alpha_1})-G_{\alpha_1}(g_{\alpha_2})\big) +G_{\alpha_2}(g_{\alpha_2})\big( G_{\alpha_1}(g_{\alpha_1 *})-G_{\alpha_1}(g_{\alpha_2 *})\big) \Big) dv_*dn \\
&+\int Bg_{\alpha_2}^{\prime } g_{\alpha_2 *}^{\prime} \Big( G_{\alpha_1}(g_{\alpha_1 *})\big( G_{\alpha_1}(g_{\alpha_2})-G_{\alpha_2}(g_{\alpha_2})\big) +G_{\alpha_2}(g_{\alpha_2})\big( G_{\alpha_1}(g_{\alpha_2 *})-G_{\alpha_2}(g_{\alpha_2 *})\big) \Big) dv_*dn \\
& -\int Bg_{\alpha_2}g_{\alpha_2 *}\Big( G_{\alpha_1}(g_{\alpha_1 *}^{\prime} )\big( G_{\alpha_1}(g_{\alpha_1}^{\prime } )-G_{\alpha_1}(g_{\alpha_2}^{\prime } )\big) +G_{\alpha_2}(g_{\alpha_2}^{\prime} )\big( G_{\alpha_1}(g_{\alpha_1*}^{\prime } )-G_{\alpha_1}(g_{\alpha_2 *}^{\prime } )\big) \Big) dv_*dn \\
&-\int Bg_{\alpha_2}g_{\alpha_2 *}\Big( G_{\alpha_1}(g_{\alpha_1*}^{\prime} )\big( G_{\alpha_1}(g_{\alpha_2}^{\prime } )-G_{\alpha_2}(g_{\alpha_2}^{\prime} )\big) +G_{\alpha_2}(g_{\alpha_2}^{\prime} )\big( G_{\alpha_1}(g_{\alpha_2 *}^{\prime} )-G_{\alpha_2}(g_{\alpha_2 *}^{\prime} )\big) \Big) dv_*dn .\hspace{.1cm} (4.8)
\end{aligned}\]
Using (\ref{T-independent-on-alpha}),
\[ \begin{aligned}
\int  B&\Big(\lvert g_{\alpha_1}g_{\alpha_1*} -g_{\alpha_2}g_{\alpha_2 *}\rvert G_{\alpha_1}(g_{\alpha_1}^{\prime} )G_{\alpha_1}(g_{\alpha_1*}^{\prime} )\Big)^\sharp dxdvdv_*dn \\
&\leq c2^{2L}\Big( \int \sup _{x\in [ 0,1] }g_{\alpha_1}^{\sharp} (t,x,v)dv
+ \int \sup_ {x\in [ 0,1] }g_{\alpha_2}^{\sharp} (t,x,v)dv\Big) \int \lvert (g_{\alpha_1} -g_{\alpha_2} )^{\sharp }(t,x,v)\rvert dxdv\\
&\leq cc_02^{2L+2}\int \lvert g^\sharp(t,x,v)\rvert dxdv.
\end{aligned}\]
We similarly obtain
\begin{eqnarray*}
\int B \Big( g_{\alpha_2}^{\prime} g_{\alpha_2 *}^{\prime} G_{\alpha_1}(g_{\alpha_1 *})\lvert ( G_{\alpha_1}(g_{\alpha_2})-G_{\alpha_2}(g_{\alpha_2})\lvert )\Big) ^\sharp dxdvdv_*dn\leq cc_02^{2L+2}|\alpha_1-\alpha_2| ,
\end{eqnarray*}
and
\begin{eqnarray*}
\int B\Big(g_{\alpha_2}g_{\alpha_2 *} G_{\alpha_1}(g_{\alpha_1 *}^{\prime} )\lvert G_{\alpha_1}(g_{\alpha_1}^{\prime} )-G_{\alpha_1}(g_{\alpha_2}^{\prime} )\rvert \Big)^\sharp dxdvdv_*dn\leq cc_02^{L+1}\int |g^\sharp(t,x,v)|dxdv.
\end{eqnarray*}
The remaining terms are estimated in the same way. It follows
\begin{eqnarray*}
\frac{d}{dt}\int|g^\sharp(t,x,v)|dxdv\leq cc_02^{2L}\Big(\int|g^\sharp (t,x,v)|dxdv+|\alpha_1-\alpha_2|\Big).
\end{eqnarray*}
 Hence
 \begin{eqnarray*}
 \lim _{(\alpha _1,\alpha _2)\rightarrow (0,0)}\sup _{t\in [0,T_L]}\int |g^\sharp(t,x,v)|dxdv= 0.
 \end{eqnarray*}
 And so $(g_\alpha )$ is a Cauchy sequence in $C([0,T_L]; L^1([0,1] \times \R ^3))$. Denote by $f$ its limit. With analogous arguments to the previous ones in the proof of this lemma, it holds that
 \begin{eqnarray*}
 \lim _{\alpha \rightarrow 0}\int \lvert Q(f)-\tilde{Q}_\alpha (g_\alpha )\rvert (t,x,v)dtdxdv= 0.
 \end{eqnarray*}
 Hence $f$ is a strong solution to (\ref{f}) on $[0,T_L]$ with initial value $f_0$. If there were two solutions, their difference denoted by $G$ would with similar arguments satisfy
 \begin{eqnarray*}
\frac{d}{dt}\int |G^\sharp(t,x,v)|dxdv\leq cc_02^{2L}\int |G^\sharp(t.x.v)|dxdv,
 \end{eqnarray*}
 hence be identically equal to its initial value zero.\\
Denote by $\mathcal{F}$ a given equibounded family of initial values bounded by $2^L$. Let $f_1$ resp. $f_2$ be the solution to (\ref{f}) with initial value $f_{10}\in \mathcal{F}$ resp. $f_{20}\in\mathcal{F}$. The equation for $\bar{g}=f_1-f_2$ can be written analogously to (4.8). Similar arguments lead to
\begin{eqnarray*}
\frac{d}{dt}\int |(f_1-f_2)^\sharp(t,x,v)|dxdv\leq cc_02^{2L}\int |(f_1-f_2)^\sharp(t,x,v)|dxdv,
 \end{eqnarray*}
 so that
 \begin{eqnarray*}
\parallel (f_1-f_2)(t,\cdot ,\cdot )\parallel _{L^1([0,1] \times \R ^3)}\leq e^{cc_0T2^{2L}}\parallel f_{10}-f_{20}\parallel _{L^1([0,1] \times \R ^3)},\quad t\in [0,T_L] .
 \end{eqnarray*}
This proves the stability statement of Theorem 1.1 .\\
If $\sup_{(x,v)\in[ 0,1] \times \R ^3}f(T_L,x,v)<2^{L+1}$, then the procedure can be repeated, i.e. the same proof can be carried out from the initial value $f(T_L)$. It leads to a maximal interval denoted by $[0,\tilde {T}_1]$ on which $f(t,\cdot ,\cdot )\leq 2^{L+1}$ and $\sup _{(t,x)\in [0,\tilde{T}_1] \times [0,1] }f(t,x,v)dv\leq c_1$ for some constant $c_1$. By induction there exists an increasing  sequence of times $(\tilde{T}_n)$  such that $f(t,\cdot ,\cdot )\leq 2^{L+n}$ on $[ 0,\tilde{T}_n]$ and $\sup _{(t,x)\in [0,\tilde{T}_n] \times [0,1] }f(t,x,v)dv\leq c_n$ for some constant $c_n$. Let $T_\infty = \lim _{n\rightarrow +\infty }\tilde{T}_n$. Either $\tilde{T}_\infty = +\infty $ and the solution $f$ is global in time, or $T_\infty $ is finite and $\overline{\lim}_{t\rightarrow T_\infty ^-}\parallel f(t)\parallel_{L^\infty ([0,1]\times \R ^3)}= +\infty$. \cqfd
%
%
%
%
\section{Conservations of mass, momentum and energy.}
\setcounter{equation}{0}
\setcounter{theorem}{0}
It follows from (\ref{T-independent-on-alpha}) that
\begin{equation}\label{control-mass-density-f}
\int \sup _{(t,x)\in [0,T_L] \times [0,1] }f(t,x,v)dv\leq 2c_0.
\end{equation}
%
%
\begin{lemma}\label{conservation-energy}
The solution $f$ to (\ref{f})) with initial datum $f_0$ conserves mass, momentum and energy.
\end{lemma}
\underline{Proof of Lemma \ref{conservation-energy}.}\\
The statement of Lemma \ref{large-velocity-2} holds for $f$, with similar proofs, using (\ref{control-mass-density-f}). Consequently,
\begin{equation}\label{large-velocities-f}
\lim _{\lambda \rightarrow +\infty }\int_{|v|\geq \lambda} \sup_{(s,x)\in [ 0,T_L] \times [0,1]}f^\#(s,x,v)dv= 0.
\end{equation}
The conservations of mass and momentum of $f$ on $[0,T_L]$ follow from the boundedness of the total energy. The energy is non-increasing by the construction of $f$. It remains to prove that the energy is non-decreasing. Taking $\psi_\epsilon=\frac{|v^2|}{1+\epsilon|v|^2}$  as approximation for $|v|^2$, it is enough to bound
\begin{eqnarray*}
\int Q(f)(t,x,v)\psi_\epsilon (v)dxdv = \int B\psi_{\epsilon}\Big( f^\prime f^\prime _{*}F(f)F(f_{*})
- ff_{*}F(f^\prime )F(f^\prime _{*})\Big) dxdvdv_*dn,\quad t\in [0,T_L],
\end{eqnarray*}
from below by zero in the limit $\epsilon \rightarrow 0$. Similarly to{ \cite{Lu2}},
\[ \begin{aligned}
\int Q(f)\psi_\epsilon dxdv
&=\frac{1}{2}\int B ff_{*}F(f^\prime )F(f^\prime _{*}\Big( \psi_\epsilon(v')+\psi_\epsilon(v'_*)-\psi_\epsilon(v)-\psi_\epsilon(v_*)
\Big)dxdvdv_*dn\\
&\geq -\int Bff_{*}F(f^\prime )F(f^\prime _{*})\frac{\epsilon |v|^2|v_*|^2}{(1+\epsilon|v|^2)(1+\epsilon|v_*|^2)}dxdvdv_*dn.
\end{aligned}\]
The previous line, with the integral taken over a bounded set in $(v,v_*)$, converges to zero when $\epsilon\rightarrow 0$. In  integrating over $|v|^2+|v_*|^2\geq2\lambda^2$ , there is symmetry between the subset of the domain with $|v|^2>\lambda^2$ and the one with $|v_*|^2>\lambda^2$. We discuss the first sub-domain, for which the integral in the last line is bounded from below by
\[ \begin{aligned}
&-c\int |v_*|^2f(t,x,v_*)dxdv_*\int_{|v|\geq \lambda} B \sup_{(s,x)\in [ 0,t] \times [0,1]}f^\#(s,x,v)dvdn\\
&\geq -c\int_{|v|\geq \lambda} \sup_{(s,x)\in [ 0,t] \times [0,1]}f^\#(s,x,v)dv,
\end{aligned}\]
which tends to zero when $\lambda \rightarrow \infty$ by (\ref{large-velocities-f}).\\
This implies that the energy is non-decreasing on $[0,T_L]$ and bounded from below by its initial value. Proving the conservation of mass, momentum and energy of $f$ on $[T_L,T_\infty [ $ is similar. That completes the proof of the lemma.     \cqfd
\\

\[\]


\begin{thebibliography}{99}
%
%
\bibitem
{AN1} L. ARKERYD, A. NOURI, {\it On a Boltzmann equation for Haldane statistics}, Preprint 2017, arXiv:1711.10357.

\bibitem
{AN2} L. ARKERYD, A. NOURI, {\it On the Cauchy problem with large data for a space-dependent anyon Boltzmann-Nordheim boson equation}, Comm. Math. Sciences, 15-5 (2017), 1247-1264.

\bibitem
{AN3} L. ARKERYD, A. NOURI, {\it Well-posedness of the Cauchy problem for a space-dependent anyon Boltzmann equation}, SIAM J. Math. Anal., 47-6 (2015), 4720-4742.




\bibitem
{B} J.-M. BONY, { \it Solutions globales born\'ees pour les mod\`eles discrets de l'\'equation de Boltzmann, en dimension 1 d'espace},  Journ\'ees "\'Equations aux d\'eriv\'ees partielles ", Exp. XVI, \'Ecole Polytech., Palaiseau, pp. 1-10, 1987.

\bibitem
{BE} M. BRIANT, A. EINAV, {\it On the Cauchy problem for the homogeneous Boltzmann-Nordheim equation for bosons: local existence, uniqueness and creation of moments}, J. Stat. Phys., 163-5 (2016), 1108-1156.

\bibitem
{CI} C.CERCIGNANI, R. ILLNER, { \it Global weak solutions of the Boltzmann equation in a slab with diffusive boundary conditions}, Arch Rat. Mech. Anal., 134 (1996), 1-16.


\bibitem
{EV2} M. ESCOBEDO, J.L. VELAZQUEZ, { \it On the blow up and condensation of supercritical solutions of the Nordheim equation for bosons}, Comm. Math. Phys., 330 (2014), 331-365.

\bibitem
{EV} M. ESCOBEDO, J.L. VELAZQUEZ, { \it Finite time blow-up and condensation for the bosonic Nordheim equation}, Invent. Math., 200 (2015), 761-847.

\bibitem
{H} F. D. HALDANE, { \it Fractional statistics in arbitrary dimensions: a generalization of the Pauli principle}, Phys. Rev. Lett., 67 (1991), 937-940.



\bibitem
{Lu1} X. LU, { \it A modified Boltzmann equation for Bose-Einstein particles: isotropic solutions and long time behaviour}, J. Stat. Phys., 98 (2000), 1335-1394.

\bibitem
{Lu2} X. LU, { \it On isotropic distributional solutions to the Boltzmann equation for Bose-Einstein particles}, J. Stat. Phys., 116 (2004), 1597-1649.

\bibitem
{Lu3} X. LU, {\it Boltzmann equation for Bose-Einstein particles: velocity concentration and convergence to equilibrium}, J. Stat. Phys.,119 (2005), 1027-1067.

\bibitem
{Lu4} X. LU,{\it The Boltzmann equation for Bose-Einstein particles; regularity and condensation}, J. Stat. Phys., 156 (2014), 493-545.

\bibitem
{N} L. W. NORDHEIM, { \it On the kinetic methods in the new statistics and its applications in the electron theory of conductivity}, Proc. Roy. Soc. London Ser. A, 119 (1928), 689-698.

\bibitem
{R} G. ROYAT, { \it Etude de l'\'equation d'Uehling-Uhlenbeck: existence de solutions proches de Planckiennes et \'etude num\'erique}, PhD, Marseille 2010.

\bibitem
{UU} E.A. UEHLING, G.E. UHLENBECK, {\it Transport phenomena in Einstein-Bose and Fermi-Dirac gases}, I. Phys. Rev. 43 (1993), 552-561

\bibitem
{V} C. VILLANI {\it A review of mathematical topics in collisional kinetic theory}, in Handboook of Mathematical Fluid Dynamics (Vol. 1), Elsevier Science (2002).

\end{thebibliography}
\end{document}